\documentclass[conference]{IEEEtran}
\IEEEoverridecommandlockouts
\usepackage{cite}
\usepackage{amsmath,amssymb,amsfonts}
\usepackage{algorithmic}
\usepackage{graphicx}
\usepackage{textcomp}
\usepackage{xcolor}

\usepackage{subfiles}
\usepackage{subcaption}
\usepackage{arydshln}
\usepackage[export]{adjustbox}

\def\ie{i.e.}
\def\eg{e.g.}
\def\etal{et al.}

\begin{document}

\title{
Efficient Feature Compression for \\ Edge-Cloud Systems 
}

\author{
\IEEEauthorblockN{Zhihao Duan}
\IEEEauthorblockA{
    \small
    \textit{Elmore Family School of Electrical and Computer Engineering} \\
    \textit{Purdue University}\\
    West Lafayette, Indiana, U.S.A. \\
    duan90@purdue.edu
}
\and
\IEEEauthorblockN{Fengqing Zhu}
\IEEEauthorblockA{
    \small
    \textit{Elmore Family School of Electrical and Computer Engineering} \\
    \textit{Purdue University}\\
    West Lafayette, Indiana, U.S.A. \\
    zhu0@purdue.edu
}
}

\maketitle

\begin{abstract}
Optimizing computation in an edge-cloud system is an important yet challenging problem. In this paper, we consider a three way trade-off between bit rate, classification accuracy, and encoding complexity in an edge-cloud image classification system. Our method includes a new training strategy and an efficient encoder architecture to improve the rate-accuracy performance. Our design can also be easily scaled according to different computation resources on the edge device, taking a step towards achieving  \textit{rate-accuracy-complexity} (RAC) trade-off. Under various settings, our feature coding system consistently outperforms previous methods in terms of the RAC performance.
Code is made publicly available\footnote{github.com/duanzhiihao/edge-cloud-rac}.
\end{abstract}


\begin{IEEEkeywords}
Feature compression, Coding for machine vision, Image classification, Edge-cloud system
\end{IEEEkeywords}

\section{Introduction}

\begin{figure}[t]
    \centering
    \includegraphics[width=0.98\linewidth]{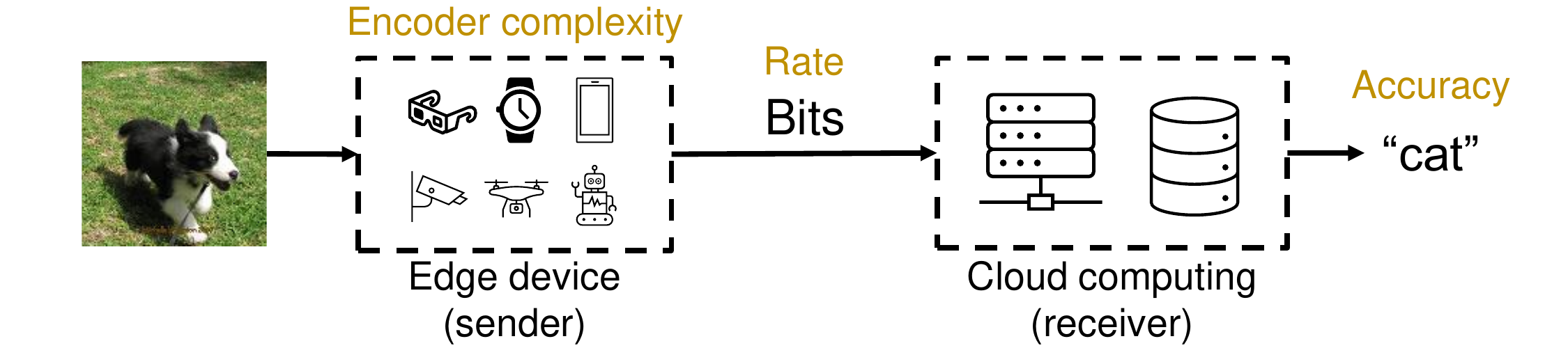}
    \caption{Illustration of an edge-cloud machine vision system, where three objectives are of interest: encoder computational complexity on the edge device, bit rate in transmission, and task accuracy in the cloud. In this paper, we present a feature compression method for the edge-cloud image classification.
    }
    \label{fig:intro_mobile_cloud}
    \vspace{-0.48cm}
\end{figure}

Mobile devices and wearable sensors are essential gateways to many autonomous systems. 
These edge devices are often deployed with vision tasks in applications involving robotics, drones, surveillance cameras, and smart home devices.
% ...
However, due to constraints on the device size and power supply, edge devices cannot support the real-time execution of neural networks-based computer vision methods, which are typically computationally expensive.
A common solution is to take advantage of the wireless connection to a powerful cloud computing resource, forming an \textit{edge-cloud system}~\cite{bajic2018icip_feature_compression, matsubara2022supervised} as illustrated in Fig.~\ref{fig:intro_mobile_cloud}.
In such a system, we obtain data (in our context, images) from edge device sensors (cameras) and send it to the cloud for task execution.

Designing practical methods for edge-cloud systems is challenging from several aspects.
First, in most cases where the edge-cloud connection is band-limited, we have constraints on the transmission bit rate, requiring the captured images to be compressed by an encoder on the edge as efficiently as possible.
Second, we also have constraints on the computational capacity of the edge device, and thus the encoder cannot be too complex.
Finally, for common vision tasks such as the image classification, we also desire high accuracy. Taking all these into consideration, we face a three-way trade-off between encoder complexity, bit rate, and vision task accuracy, which We term as the rate-accuracy-complexity (RAC) trade-off.

A straightforward baseline for RAC trade-off is to compress and transmit images using different codecs.
However, image codecs are typically developed for image reconstruction, making them suboptimal for vision task execution.
Recently, research focused on feature compression has gained attention, which can be optimized for different vision tasks~\cite{bajic2018icip_feature_compression, singh2020icip_compressible_feature}.
Another advantage to transmit features rather than images is privacy protection. By transmitting features, one could skip pixel reconstruction on the cloud and thus reduce potentially exposing image contents to the attackers~\cite{kakkad2019biometric}.

Entropic Student~\cite{matsubara2022supervised} is a recently proposed feature compression framework that provides promising rate-accuracy performance. However, it does not take into account the encoder complexity as part of the performance trade-off.
To extend the scope to the three-way RAC trade-off, we propose a flexible encoder whose complexity can be easily adjusted for different device capacities. Together with a new training strategy, we achieve better RAC performance compared to previous methods, including image coding baselines and the state-of-the-art Entropic Student model.

Our contributions can be summarized as follows:
\begin{itemize}
    \item We expand the scope of previous feature compression systems to include the RAC trade-off.
    \item We design a flexible neural network architecture for feature coding, which is both efficient and achieves good classification performance.
    \item We conduct extensive experiments to verify the effectiveness of our approach, showing its advantages over previous methods.
\end{itemize}

\section{Related Work}
\subsection{Image Compression}
\textbf{Traditional image coders}.
Traditional image codecs typically follow a linear transform coding paradigm, such as block-based discrete cosine transform (DCT) and discrete wavelet transform (DWT).
Representative codecs include JPEG, WebP, and BPG (\ie, HEVC intra~\cite{lainema2012hevc_intra}).
Their encoding complexity varies depending on different factors such as the choice of the entropy coding algorithms, number of prediction modes, and how much optimization is done during training.
In general, more complex computation devoted at the encoder leads to better compression performance.

\textbf{Learned image coders}.
With the rise of deep learning, learned image coders (LICs) extend the traditional linear transform coding paradigm to a non-linear fashion, which is achieved by transforming images using neural networks~\cite{balle2016end2end}.
Despite improved compression performance compared to traditional codecs, LICs typically require more computational resources due to highly parameterized neural network layers and autoregressive entropy models~\cite{minnen2018joint}.

The paradigm of image coding share a common disadvantage: they are designed for image reconstruction and may be suboptimal for vision tasks in most edge-cloud settings.

\subsection{Feature compression}
Various methods have been proposed to compress and transmit features rather than images for machine vision.

\textbf{Hand-crafted feature coding}.
Earlier works have explored compressing deep features using hand-crafted algorithms. Choi and Bajić use a block-based intra and inter coding algorithm for video feature compression in~\cite{bajic2018mmsp} and a HEVC-based feature compression method for object detection in~\cite{bajic2018icip}. Following their works, methods have been proposed for feature compression using pruning~\cite{bajic2020back_and_force}, quantization~\cite{bajic2020icme}, and bidirectional prediction~\cite{bajic2020back_and_force}.


\textbf{Learned feature compression}.
Instead of manually designed methods, recent work has shifted attention to end-to-end \textit{learned} feature compression and showed promising results.
Singh \etal{} use a fully factorized entropy model to constrain the entropy of features, which results in improved rate-accuracy trade-off over previous methods~\cite{singh2020icip_compressible_feature}. Dubois \etal{} show that, given enough computational capacity, features can be extremely (around $1,000\times$) compressed without affecting the vision task accuracy~\cite{dubois2021nips_lossylossless}.
More recently, Matsubara \etal{} train vision models with feature compressor jointly and show state-of-the-art rate-accuracy performance~\cite{matsubara2022supervised}.

While existing methods mostly focus on improving the rate-accuracy performance, their approach lacks scalability for different encoder complexity. In this work, we extend the scope to rate-accuracy-complexity (RAC) trade-off by proposing a easily scalable encoder architecture, which can be adapted to edge devices with different computation capacities.


\section{Preliminaries}
\label{sec:background}

\begin{figure}[t]
    \centering
    \begin{subfigure}[b]{\linewidth}
        \centering
        \includegraphics[width=0.96\linewidth]{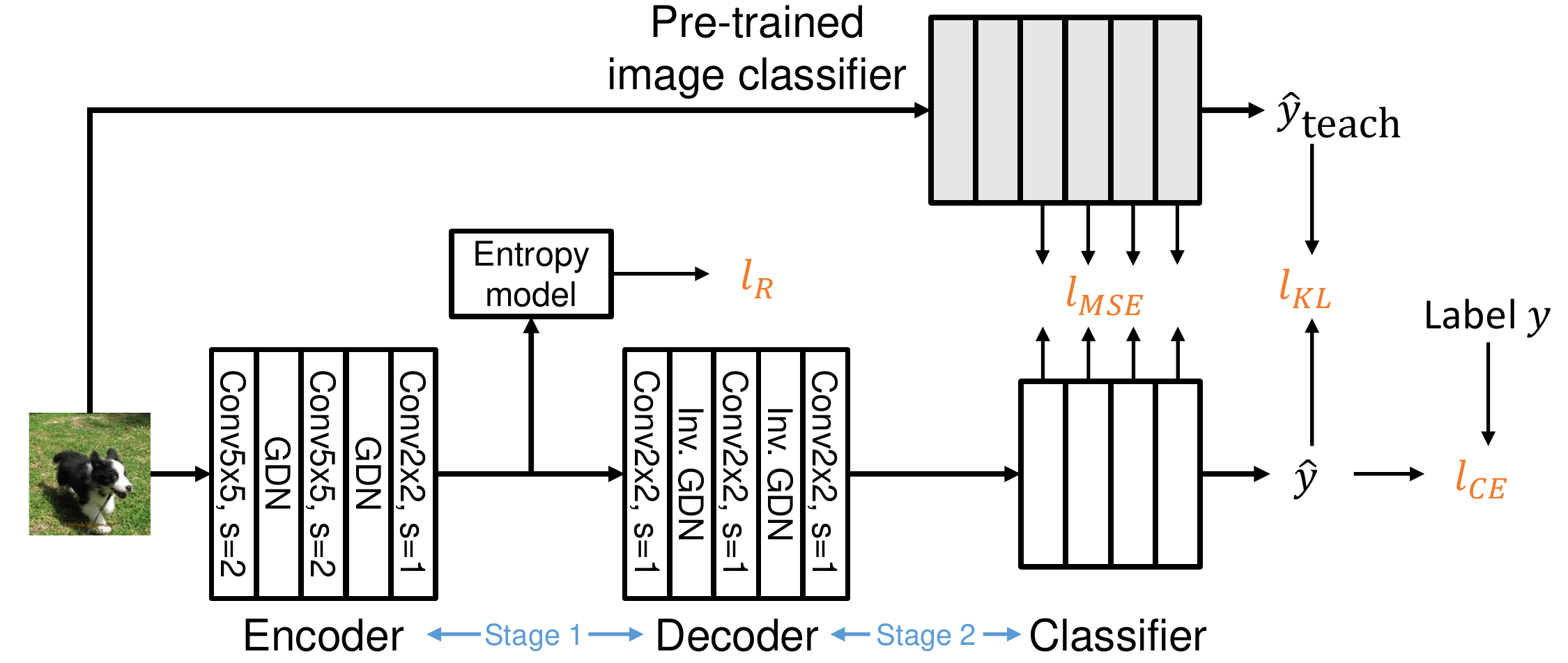}
        \vspace{-0.1cm}
        \caption{Training overview. GDN refers to the generalized divisible normalization layer~\cite{balle2016gdn}, the entropy model is fully factorized, and the classifier is ResNet-50~\cite{he2016resnet}. See text for description.}
        \label{fig:background_sc2_training}
    \end{subfigure}
    \hfill
    \vspace{-0.16cm}
    \begin{subfigure}[b]{\linewidth}
        \centering
        \includegraphics[width=0.86\linewidth]{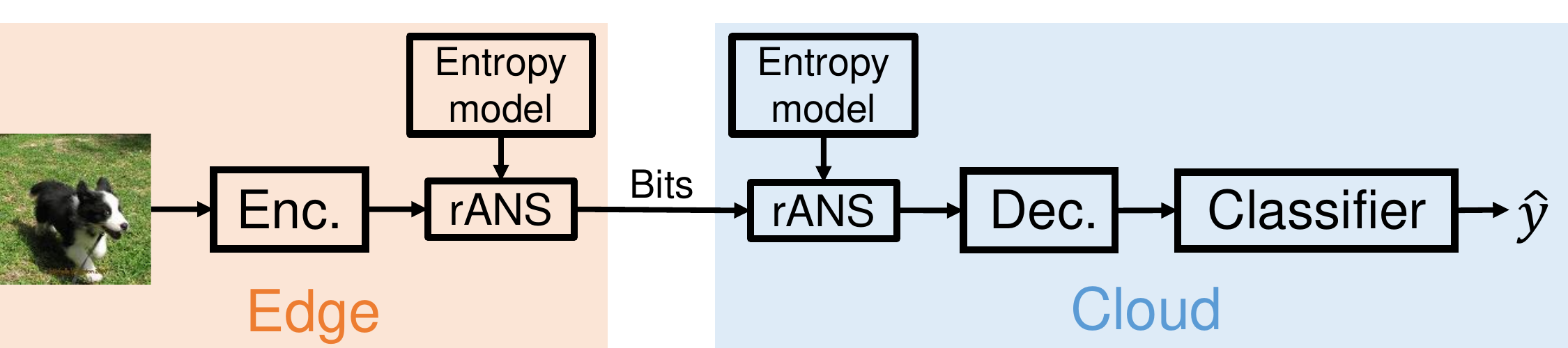}
        \caption{Testing (deployment). Entropy coding is done by the range-based variant of asymmetric numeral systems (rANS)~\cite{duda2013ans}.}
        \label{fig:background_sc2_testing}
    \end{subfigure}
    \hfill
    \vspace{-0.48cm}
    \caption{Model training (a) and deployment (b) in the Entropic Student framework~\cite{matsubara2022supervised} for feature compression.}
    \label{fig:background_sc2}
\end{figure}

The current best method in terms of rate-accuracy performance, Entropic Student~\cite{matsubara2022supervised}, is a good starting point for our proposed work.
In this section, we briefly summarize it to provide the background of our method, and we refer interested readers to the original paper~\cite{matsubara2022supervised} for more details.

The framework for classification is shown in Fig.~\ref{fig:background_sc2}.
For split computing, a feature encoder is designed to be deployed on the edge device, and a decoder together with a classifier are executed on the cloud.
Specifically, a ResNet-50~\cite{he2016resnet} classifier is adopted for image classification.

Training is done in two stages.
In the first stage, the classifier is initialized using a pre-trained model and fixed (\ie, no parameter updating), and the encoder and decoder are learned to compress images and reconstruct desired features.
In the second stage, the encoder is fixed, while the decoder and the classifier are trained to optimize the task accuracy.
The training loss functions are:
\begin{equation}
\begin{aligned}
L_\text{stage1} &= \lambda l_{R} + l_\mathit{MSE},
\\
L_\text{stage2} &= \lambda l_{R} + 0.5 \cdot (l_\mathit{KL} + l_\mathit{CE}),
\end{aligned}
\end{equation}
where $l_{R}$ is the bit rate loss estimated by a factorized entropy model~\cite{balle2016end2end}, $l_\mathit{MSE}$ is the mean squared error (MSE) between the classifier's intermediate features and the ones from a pre-trained teacher model, $l_\mathit{KL}$ is the KL divergence loss for knowledge distillation~\cite{hinton2015kd}, $l_\mathit{CE}$ is the standard cross-entropy loss for classification, and $\lambda$ is a scalar hyperparameter to trade-off between rate and classification accuracy.

The deployment scenario is shown in Fig.~\ref{fig:background_sc2_testing}, where the entropy model is shared between the sender (edge) and the receiver (cloud) for entropy coding, implemented by the range-based asymmetric numeral system (rANS) algorithm~\cite{duda2013ans}.

Despite its strong rate-accuracy performance~\cite{matsubara2022supervised}, the Entropic Student model only operates at a single encoder complexity and is not feasible for RAC trade-off.
In the next section, we introduce our method, which not only improves the rate-accuracy performance, but also enables the RAC trade-off.


\begin{figure*}[t]
    \begin{subfigure}[b]{0.36\linewidth}
        \centering
        \includegraphics[width=\linewidth]{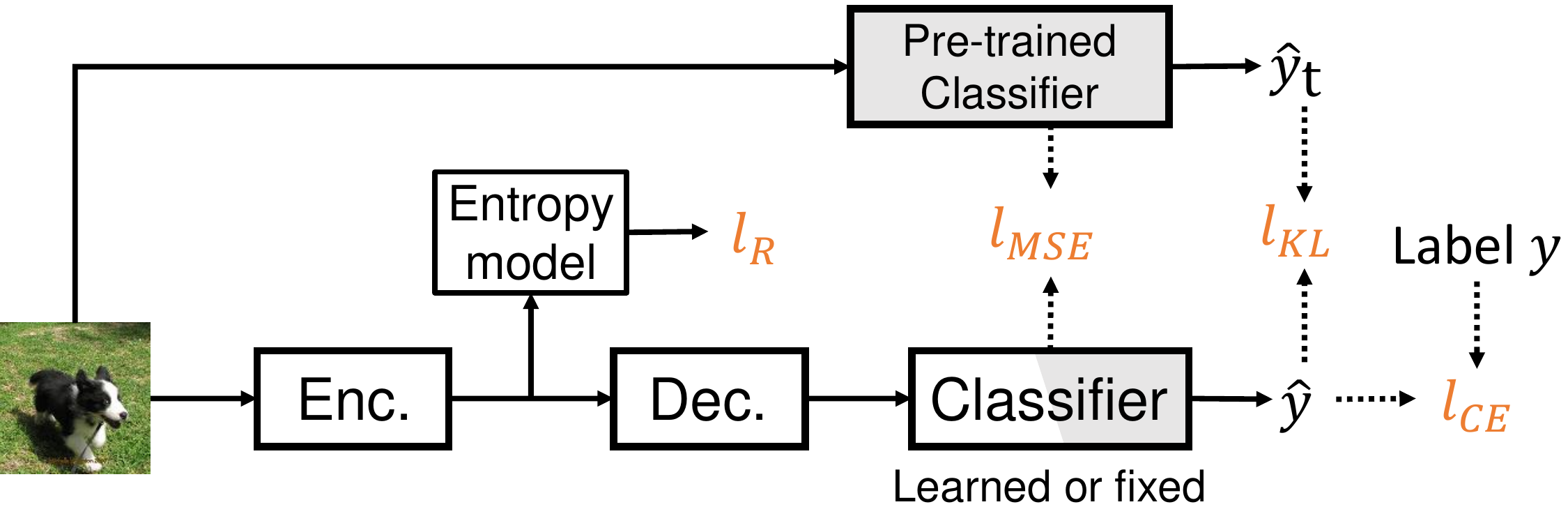}
        \caption{Training overview.}
        \label{fig:method_ours_training}
    \end{subfigure}
    \hfill
    \begin{subfigure}[b]{0.22\linewidth}
        \centering
        \includegraphics[width=\linewidth]{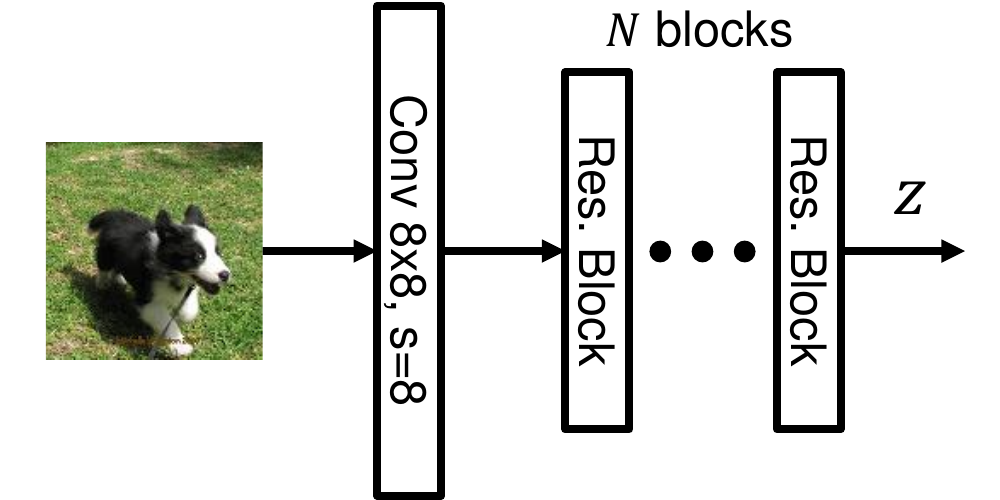}
        \caption{Encoder architecture.}
        \label{fig:method_ours_encoder}
    \end{subfigure}
    \hfill
    \begin{subfigure}[b]{0.14\linewidth}
        \centering
        \includegraphics[width=\linewidth]{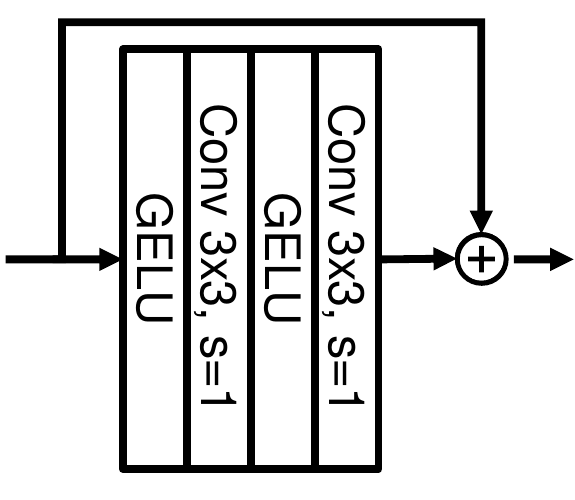}
        \caption{Residual block.}
        \label{fig:method_ours_res}
    \end{subfigure}
    \hfill
    \begin{subfigure}[b]{0.22\linewidth}
        \centering
        \includegraphics[width=\linewidth]{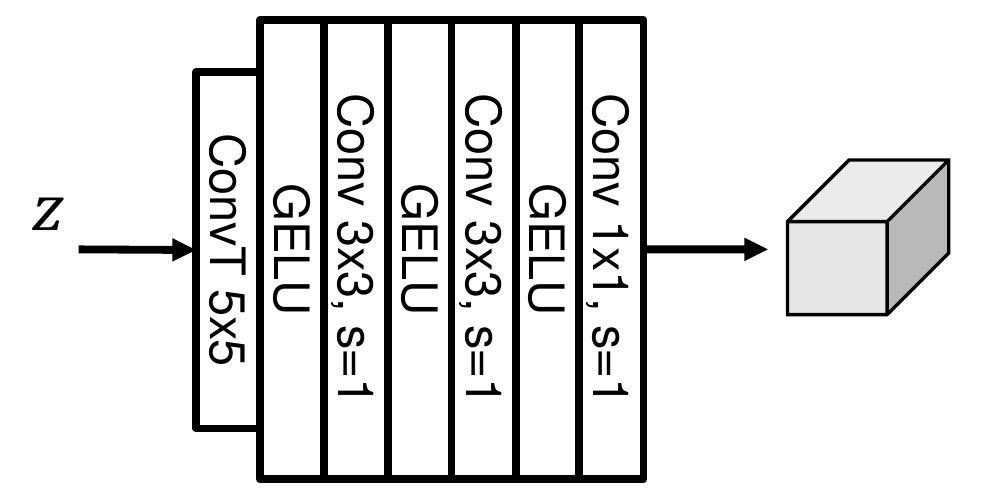}
        \caption{Decoder architecture.}
        \label{fig:method_ours_decoder}
    \end{subfigure}
    \hfill
    \caption{Summary of proposed method. In (a), we show the training strategy, where the entropy model is fully factorized, and gray blocks indicate frozen models (\ie, no parameter update during training). The classifier can either be trained jointly with the system, or initialized with a pre-trained model and fixed throughout training.
    In (b) and (c), we show our encoder and decoder architecture, respectively. Our encoder network complexity can be tuned by adjusting the number of residual blocks, $N$.
    For the classifier, we use a truncated version of ResNet-50, following~\cite{matsubara2022supervised}.
    }
    \label{fig:method_ours}
\end{figure*}

\section{Our Approach}

Inspired by the success of knowledge distillation in the Entropic Student model, we adopt a similar framework and loss functions, shown in Fig.~\ref{fig:method_ours_training}.
However, we make modifications including a new training strategy and network architecture,
by which we are able to realize the RAC trade-off.

\subsection{Single-stage Training}
The Entropic Student method adopts a two-stage training strategy, where the encoder and decoder are trained in the first stage, and the decoder and classifier are fine-tuned in the second stage.
We note that this decoupled training strategy requires more hyperparameter searching, and it is also sub-optimal for rate-accuracy performance since network components are trained separately.
Instead, we propose a simple fix, \ie, to train the encoder, decoder, and classifier in an end-to-end fashion.
Our single-stage loss function is as follows:
\begin{equation}
\label{eq:method_ours_loss}
L_\text{single} = \lambda l_{R} + l_\mathit{MSE} + 0.5 \cdot (l_\mathit{KL} + l_\mathit{CE}),
\end{equation}
where the terms are defined in Sec.~\ref{sec:background}. We keep the same scaling factor for each loss term as in the Entropic Student without further tuning.
Later in Sec.~\ref{sec:exp_ablation}, we show that our single-stage strategy improves the rate-accuracy trade-off despite being simpler to implement.

\subsection{Network Architecture} \label{sec:method_architecture}
We introduce an encoder network architecture specifically designed for efficient and flexible feature compression, shown in Fig.~\ref{fig:method_ours_encoder}.
We now describe our design choices in details.

\textbf{Downsampling layer.} We first noticed that the 5x5 convolutions for downsampling in previous works (\eg, in the Entropic Student~\cite{matsubara2022supervised} and learned image coders~\cite{balle2016end2end, balle18hyperprior}) are computationally inefficient due to the large overlap between convolutional kernels.
To remove this overlap, we adopt a single \textit{patch embedding}~\cite{dosovitskiy2021vit} layer for downsampling, which can be equivalently viewed as a convolution with kernel size equal to its strides.

\textbf{Network blocks.} Instead of feedforward convolutions as in previous works, we adopt residual blocks in our feature encoder, which enables easy adjustment of the encoder complexity simply by setting the number of residual blocks, $N$.
A larger $N$ (\ie, more residual blocks in the encoder) improves the rate-accuracy performance while results in higher encoder complexity. Thus by tuning $N$, we can trade-off between the rate-accuracy performance and the encoding complexity of our system.
We use a standard design for the residual blocks as shown in Fig.~\ref{fig:method_ours_res}.
For non-linearity layers, we use the GeLU activation~\cite{hendrycks2016gelu} instead of GDN~\cite{balle2016gdn}, as we empirically found that the former leads to better performance with the same computational complexity.

\textbf{Downsampling ratio.}
Our final step is to use $8\times$ downsampling, instead of the $4\times$ downsampling in Entropic Student, during encoding.
Our choice is motivated by the observation that $8\times$ downsampling is a common choice for various image processing tasks (\eg, in traditional codecs like JPEG) and computer vision tasks (\eg, in image object detection~\cite{redmon2018yolov3} and semantic segmentation~\cite{zhao2017psp}).
Note that in our feature decoder on the cloud (Fig.~\ref{fig:method_ours_decoder}), we upsample the feature by $2\times$ to recover spatial details of the original image.
With this configuration of the encoder and the decoder, we are able to reduce the encoding latency and improve the average accuracy by a large margin, as discussed later in Sec.~\ref{sec:exp_ablation}.

\subsection{Feature Compression for a Fixed Classifier}
\label{sec:method_cls_fixed}
So far, we have considered the \textit{joint} training setting, where the classifier and our feature coder are trained jointly.
In addition, we also examine the setting in which a pre-trained classifier is available but fixed, illustrated in Fig.~\ref{fig:method_ours_training}.
This scenario may happen in practice where there are different edge devices sharing a single cloud vision model.
The edge devices may use different feature encoders/decoders according to their computational capacity, but the vision model on the cloud should be fixed.
In this case, we use the same loss function as in Eq.~\ref{eq:method_ours_loss}, but we train the encoder and decoder networks without updating the classifier.
We denote the methods in this setting by \textit{cls. fixed} to distinguish them from the \textit{joint} training setting, where the classifier can be trained together with the encoder and the decoder.

\section{Experiments}


We compare our method to previous ones for edge-cloud image classification. For fair comparison, all methods use the ResNet-50~\cite{he2016resnet} (for image coding methods) or a truncated version of ResNet-50 (for feature coding methods) as the classifier.

\subsection{Dataset and Metrics}
\label{sec:exp_metrics}
We use the 1,000-class ImageNet dataset~\cite{li2009imagenet} for training and testing.
Standard data augmentations, including random resizing and cropping, are applied during training.
We use the ImageNet \textit{val} set for testing, in which we resize and center-crop all images such that they have a resolution of $224 \times 224$.

All metrics are computed on the $224 \times 224$ \textit{val} set.
Bit rate is measured in terms of bits per pixel (bpp).
For classification accuracy, we report the top-1 accuracy of the classifier predictions.
For encoding complexity, we focus on the CPU latency since it well simulates real-world scenarios.
All encoding latencies are measured using an Intel 12700K CPU, averaged over the first 5,000 images in the \textit{val} set.


We also introduce a new metric, Delta-accuracy, which computes the average accuracy improvement (w.r.t. a baseline method) over all bit rates. By using this metric, we can summarize each rate-accuracy curve into a scalar, thus being able to represent the three-way RAC trade-off in a 2-D plot.
Our computation of Delta-accuracy follows the convention of BD-PSNR and BD-rate~\cite{bjontegaard2001bdrate} except one aspect.
Unlike PSNR, the unit of accuracy is percentage (not decibel), so we compute average accuracy improvement over the \textit{normal} bit rate space (instead of the logarithmic rate scale used in BD-PSNR). 

\subsection{Experiment Settings}
\label{sec:exp_baselines}
We consider the following baseline methods.

\textbf{Image coding baselines.} We compress images using an image codec at the sender side, and we reconstruct images at the receiver side using the same codec.
Then, we run ResNet-50 on the reconstructed images to infer class prediction.
Our selected codecs include handcrafted (\eg, WebP and BPG~\cite{lainema2012hevc_intra}) and learning-based methods (\eg, the Factorized model~\cite{balle2016end2end}, and the Hyperprior model~\cite{balle18hyperprior}).

\textbf{Feature coding baseline.} We use the existing best method, Entropic Student~\cite{matsubara2022supervised}, as a powerful baseline.

As we discussed in Sec.~\ref{sec:method_cls_fixed}, we group methods into two groups so that methods within each can be fairly compared.

\textbf{Classifier fixed setting:} The pre-trained ResNet-50 is used and remains fixed (\ie, no fine-tuning).
The image coding baselines all fall in this setting. We label the methods in this setting as \textit{cls. fixed}.

\textbf{Joint training setting:} The pre-trained ResNet-50 is jointly trained with the feature encoder and/or decoder. The Entropic Student model falls in this setting. We label the methods in this setting as \textit{joint}.



\subsection{Main Results for RAC Trade-off}
\begin{figure}[t]
    \centering
    \includegraphics[width=0.86\linewidth]{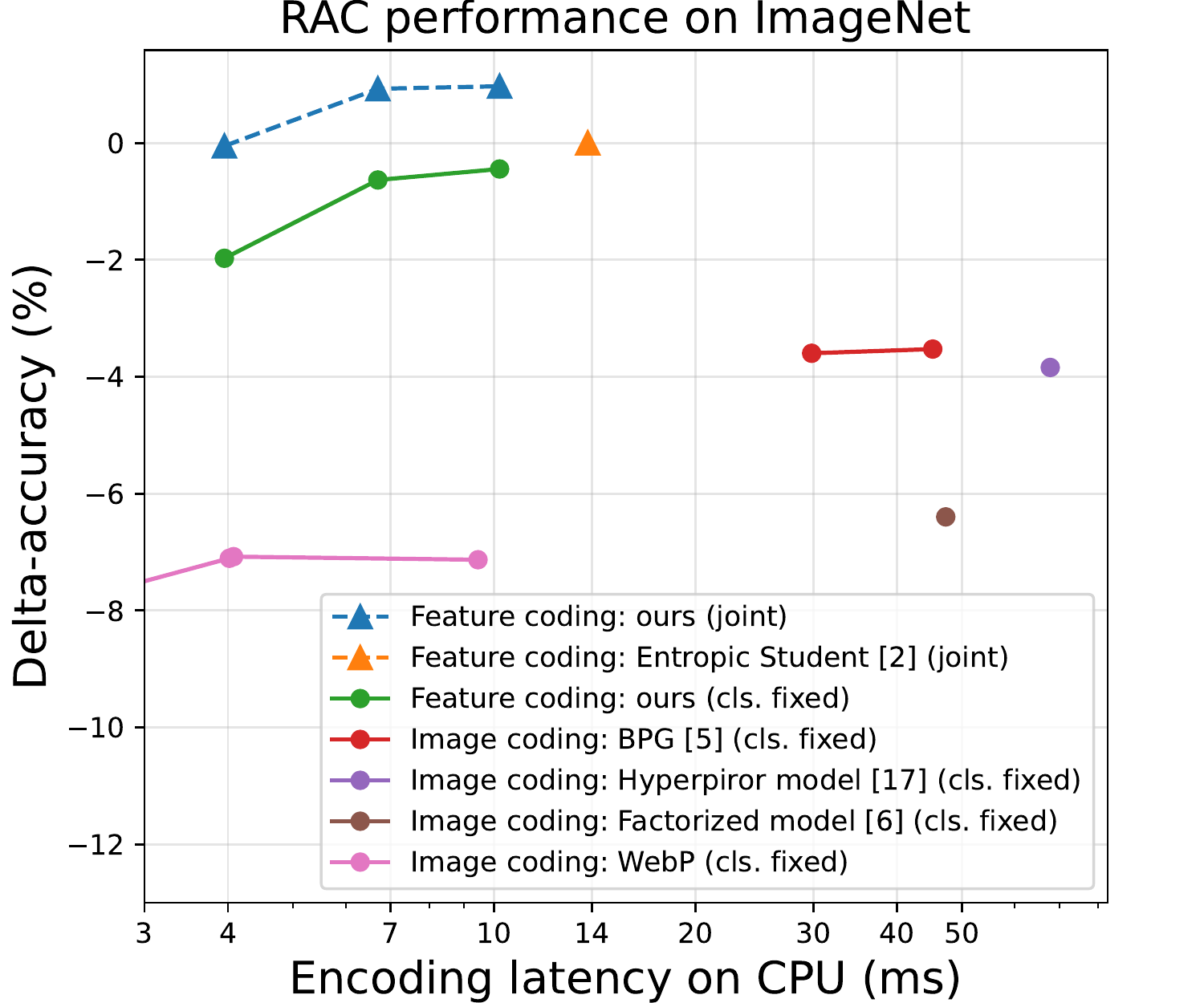}
    \caption{RAC trade-off, visualized by Delta-accuracy vs encoding complexity. The Delta-accuracy is w.r.t. the Entropic Student baseline.}
    \vspace{-0.24cm}
    \label{fig:exp_rac}
\end{figure}

To benchmark our method against the baselines in terms of RAC performance, we first measure the rate-accuracy curves for all baselines and our methods. Then, we compute the Delta-accuracy of each method w.r.t. the Entropic Student baseline (whose Delta-accuracy is 0 by definition).
We show the Delta-accuracy as a function of encoding latency in Fig.~\ref{fig:exp_rac}.
Note that the Entropic Student~\cite{matsubara2022supervised} corresponds to a single point on the complexity vs. Delta-accuracy plot since its complexity cannot be tuned. Similarly, the Factorized model~\cite{balle2016end2end} and Hyperprior model~\cite{balle18hyperprior} also produce a single point on the this plot.
For WebP and BPG~\cite{lainema2012hevc_intra} codec, we control the encoding latency by setting the compression speed option provided by the codec implementation.
For our method, we control the latency by choosing $N \in \{0,4,8\}$, where $N$ is the number of residual blocks in the encoder.

From Fig.~\ref{fig:exp_rac}, we observe that for \textit{classifier fixed} methods, our approach clearly outperforms the image coding baselines by a large margin, demonstrating the effectiveness of feature coding.
This improvement is expected, because our feature encoder and decoder are explicitly trained to maximize classification accuracy, instead of pixel reconstruction as in image coding baselines.
For the \textit{joint training} setting, our method achieves higher Delta-accuracy than the Entropic Student baseline with approximately only half of the encoding latency.
In addition, the flexible design of our encoder enables encoding complexity adjustment, as opposed to the Entropic Student, by which only a single complexity can be achieved.

\subsection{Ablation Study} \label{sec:exp_ablation}
To analyze how each component of our method contributes to the RAC improvement, we start from the Entropic Student baseline, and we include our proposed components one at a time. At each step, we show the change of rate-accuracy performance as well as the encoding complexity.

\begin{figure}[t]
    \centering
    \includegraphics[width=0.92\linewidth]{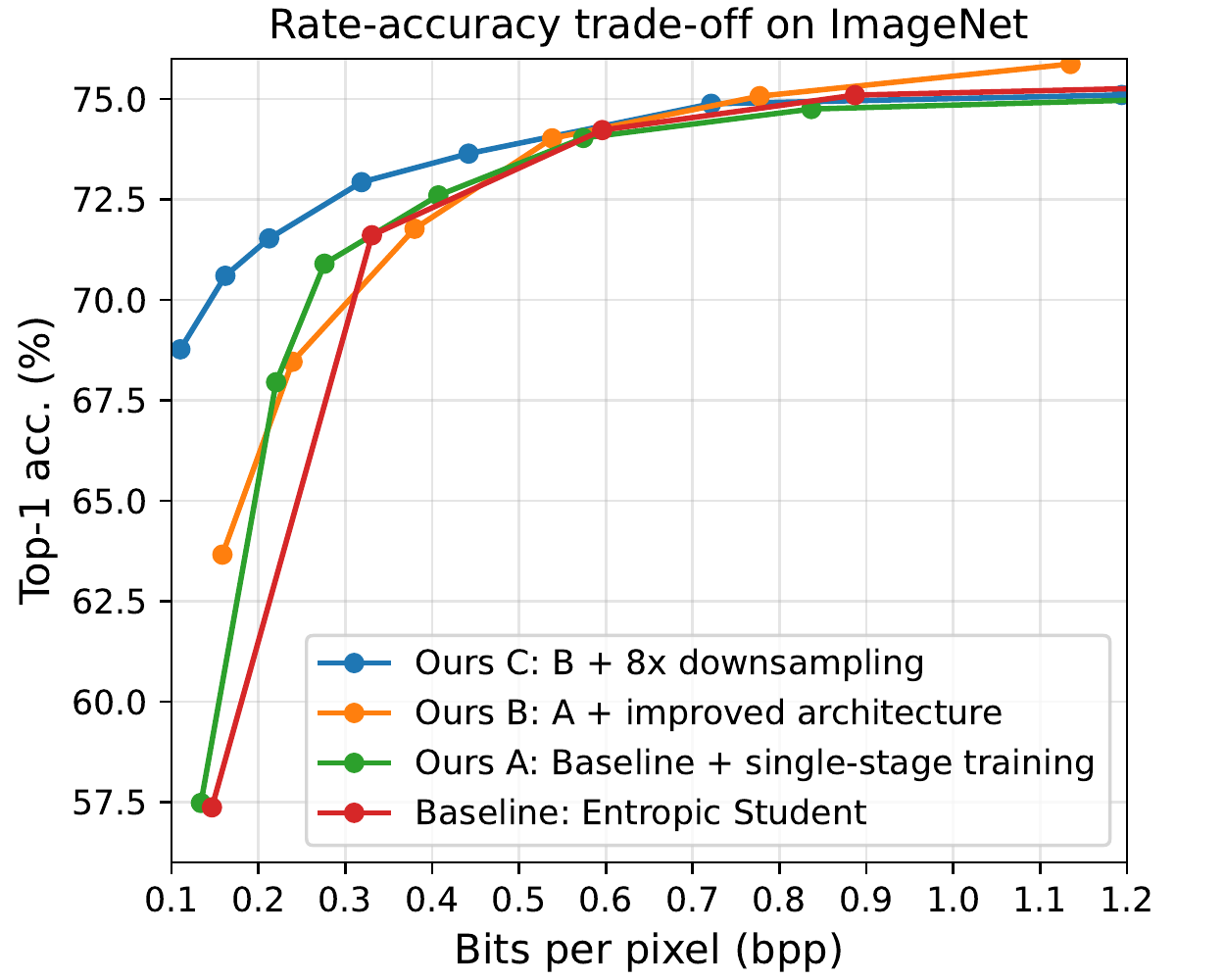}
    \vspace{-0.16cm}
    \caption{Ablation study of our approach compared to the Entropic Student baseline. Trained and tested on ImageNet~\cite{li2009imagenet}.}
    \label{fig:method_ablation}
\end{figure}

We first compare our single-stage strategy (referred to as \textit{configuration A}) with the original two-stage one in Fig.~\ref{fig:method_ablation}. In Table~\ref{table:method_configuration}, we also show the encoding complexity and Delta-accuracy.
We observe that the single-stage training does not impair the performance; instead, the average accuracy slightly increases by $0.13\%$ with encoding complexity unchanged.

Then, we replace the Entropic Student encoder by our patch embedding layer and residual blocks, and we show the resulting system as \textit{configuration B} in Fig.~\ref{fig:method_ablation} and Table~\ref{table:method_configuration}. Note that the downsampling ratio ($4 \times$) is unchanged.
We can see that the improved architecture reduces the encoding latency by more than 30\% and improves the average accuracy by an additional small percentage ($0.07\%$).

Finally, we increase the downsampling ratio to $8\times$ as we discussed in Sec.~\ref{sec:method_architecture}, and we refer to this version as \textit{configuration C}.
By increasing the downsampling ratio, we again reduce the encoding latency and improve the average accuracy by an additional large margin, after which we achieve an overall latency reduction of $50\%$ and an average accuracy improvement of $0.93\%$.
From Fig.~\ref{fig:method_ablation}, we notice that the improvement is most significant at lower bit rates.
We conjecture that lower bit rates naturally call for a more compact compressed representation, which can be achieved by more aggressive downsampling.
For higher bit rate (\eg, more than 0.8 bpp), our approach remains on par with the baseline.

\begin{table}[t]
\centering
\caption{
Ablation study. FLOPs and encoding latency are for 224x224 resolution inputs.
All percentages are w.r.t. to the baseline, Entropic Student.
}
\vspace{-0.16cm}
\begin{adjustbox}{width=1\linewidth}
\begin{tabular}{|l|cc|c|}
\hline
                            & \textbf{FLOPs $\downarrow$} & \textbf{Enc. Latency $\downarrow$} & \textbf{Delta-acc.$\uparrow$}   \\ \hline
Baseline: Entropic Student  & 0.466B         & 13.8ms           & 0.0\%                        \\ \hline
A: Baseline + single-stage  & 0.466B (-0.0\%)& 13.8ms (-0.0\%)  & +0.13\%                       \\
B: A + improved architecture & 0.457B (-0.02\%)& 9.10ms (-34.1\%) & +0.20\%                       \\
C: B + 8x downsampling      & 0.489B (+0.05\%) & 6.70ms (-51.4\%) & +0.93\%              \\ \hline
\end{tabular}
\end{adjustbox}
\label{table:method_configuration}
\end{table}




\section{Conclusion}
In this paper, we present a feature coding system for image classification in edge-cloud systems.
We show that by proper design of the network architecture, existing feature coding methods can be improved and extended to consider the RAC trade-off.
Our system outperforms all previous methods in terms of RAC trade-off, which we believe is valuable for many real-world applications deployment.
A limitation of our system is the requirement of a separate neural network model for every rate-complexity pair, which we leave to our future work to solve.


{
\bibliographystyle{IEEEtran.bst}
\bibliography{references.bib, mobilecloud.bib}
}

\end{document}